\pdfoutput=1
\documentclass[12pt]{article}
\usepackage{graphicx}
\usepackage{float} 

\begin{document}
\begin{titlepage}

\title{The Work-Energy Relation for Particles on Geodesics in the pp-Wave 
Spacetimes}

\author{J. W. Maluf$\,^{(1)}$, J. F. da Rocha-Neto$\,^{(2)}$, \\
S. C. Ulhoa$\,^{(3)}$, and F. L. Carneiro$\,^{(4)}$ \\
Instituto de F\'{\i}sica, 
Universidade de Bras\'{\i}lia\\
70.919-970 Bras\'{\i}lia DF, Brazil\\}
\maketitle
\bigskip
\bigskip

\begin{abstract}
A non-linear gravitational wave imparts gravitational acceleration to all 
particles that are hit by the wave. We evaluate this acceleration for particles
in the pp-wave space-times, and integrate it numerically along the
geodesic trajectories of the particles during the passage of a burst of 
gravitational wave. The time dependence of the wave is given by a Gaussian,
so that the particles are free before and after the passage of the wave. The
gravitational acceleration is understood from the point of view of a flat 
space-time, which is the initial and final gravitational field configuration.
The integral of the acceleration along the geodesics is the analogue
of the Newtonian concept of work per unit mass. Surprisingly, it yields 
almost exactly the variation of the non-relativistic kinetic energy per unit 
mass of the free particle. Therefore, the work-energy relation 
$\Delta K = \Delta W$ of classical Newtonian physics also holds for a particle
on geodesics in the pp-wave space-times, in a very good approximation, and 
explains why the final kinetic energy of the particle may be smaller or larger 
than the initial kinetic energy. 
\end{abstract}
\thispagestyle{empty}
\bigskip
%\noindent {\bf Keywords}: Non-linear gravitational waves; Gravitational energy 
%transfer; Variations of the kinetic energy of particles; Free particles and 
%gravitational plane waves.\par
\vfill
\begin{footnotesize}
\par
\end{footnotesize}

\bigskip
{\footnotesize
\noindent (1) wadih@unb.br, jwmaluf@gmail.com\par
\noindent (2) rocha@fis.unb.br\par
\noindent (3) sc.ulhoa@gmail.com\par
\noindent (4) fernandolessa45@gmail.com\par}

\end{titlepage}
\newpage

\section{Introduction}

Non-linear gravitational waves are exact solutions of Einstein's equations that
represent the propagation of non-trivial configurations of the gravitational 
field. Assuming that the time dependence of the wave is modelled by a Gaussian
function, the region within the propagating Gaussian is endowed with a 
gravitational field that exhibits interesting properties. Recent investigations
of these phenomena have led to analyses of the memory effect 
\cite{ZDGH1,ZDGH2,ZDGH3,ZDGH4,KASP}. The memory effect may be understood as the
permanent displacement in the detector (a collection of free particles, for
instance) after the passage of a gravitational wave. The idea was first put 
forward by Zel´dovich and Polnarev \cite{ZP}, and later by Braginsky and 
Grishchuk \cite{BG}. The limit in which the Gaussian width tends to zero, and 
the Gaussian tends to a delta function, yields impulsive gravitational waves, 
which have also been thoroughly investigated 
\cite{Steinbauer,Podolsky-2,Saemann,Podolsky-3}. Recently, as an interesting
attempt to unveil the features of the memory efect, it has been suggested
\cite{ZEGH} that a matrix Sturm-Liouville equation, constructed out of the field
quantities of the wave, plays a central role in the determination of the 
trajectories of the particles that are initially at rest.

In the analysis of the trajectories and velocities of free particles in the 
presence of pp-waves (the particles are understood to be free before and after 
the passage of the wave), we have found that the final kinetic energy of the 
free particles may be smaller or larger than the initial values
\cite{Artigo1,Artigo2}, and also that the final angular momentum of the particle
may be smaller or larger in magnitude than the initial values \cite{Artigo3}. 
At first sight it might sound strange that the particle loses kinetic energy,
most probably by transferring this energy to the wave. We have found, however, a
very simple explanation to this phenomenon. The idea is based on the classical
Torricelli equation, $v^2_f=v^2_i+2a\Delta x$, that generalizes to the 
expression

\begin{equation}
{1\over 2} v^2_f-{1\over 2}v^2_i = \int^f_i{\bf a}\cdot d{\bf l}\,,
\label{1}
\end{equation}
for a particle that undergoes an acceleration ${\bf a}$ between the initial and
final positions. By multiplying the equation above by the particle mass $m$,
we obtain the standard work-energy relation $\Delta K = \Delta W$ of classical
Newtonian physics, for a particle that is under the action of a force 
${\bf F}=m{\bf a}$.

In this article we will identify ${\bf a}$ with the gravitational 
acceleration, establish the right hand side of Eq. (\ref{1}) in 
relativistic form, and verify that the resulting equation is satisfied, to an
excellent approximation, along the trajectory of a particle in the pp-wave
space-time, by evaluating numerically both sides of the equation. Therefore, we
conclude that the acceleration due to the gravitational wave is responsible for
the variation of the kinetic energy of the particle, and is ultimately relevant
to the analysis of the memory effect. The analytical expression of $\Delta W$ 
represents the precise amount of energy per unit mass that is exchanged between
the gravitational field and the particle.

Although the equations that describe the pp-waves are well known, the intrinsic 
physical features of the waves are not completely clear. In order to gain 
further insight on the pp-waves and to analyse them from alternative points of 
view, we construct the centre of mass density of the gravitational  
wave using the definition of centre of mass that arises in the 
teleparallel equivalent of general relativity (TEGR). We believe that there is
a close relationship between the gravitational acceleration in a certain 
space-time and the density of centre of mass of the gravitational field. 
According to previous investigations, the gravitational acceleration that acts
on geodesic particles is directed toward regions of higher intensity of the 
gravitational centre of mass, described in this article by $M$. If $M>0$, as we
will see, then necessarily the radial gravitational acceleration $a_{\rho}$
is attractive,  i.e., $a_{\rho}<0$.

The article is organised as follows. In section 2, we present the equations for
the pp-waves in suitable coordinates, and construct the set of tetrad fields 
that establish a frame adapted to static observers in the space-time. Then we 
review the construction of the acceleration tensor on frames. This tensor 
determines the inertial accelerations that are necessary to maintain the frame
in a certain inertial state (i.e., to maintain the frame static in space-time,
for instance). This tensor 
is evaluated for the set of tetrad fields adapted to a static observer. The 
gravitational acceleration on the frame is precisely minus the inertial
acceleration, and is identified with the acceleration that acts on the otherwise
free particles. In section 3 we construct the generalised expression for the
work-energy relation in space-time, and obtain the main results of the article.
In section 4 we present a very brief exposition of the TEGR, and recall that
the energy-momentum and 4-angular momentum of the gravitational field satisfy
the algebra of the Poincar\'e group in the phase space of the theory. Then, we
evaluate the centre of mass density of the gravitational wave, and show that
the above mentioned relationship between the centre of mass density and the
gravitational acceleration takes place in the context of the gravitational field
of the wave. In section 5 we consider the interesting case of an
oscillating pp-wave, where we explicitly introduce a harmonic function in the
function $H$ that determines the wave. Finally, in section 6 we present the 
conclusions.

\section{Plane-fronted gravitational waves and static frames in space-time}

The mathematical construction of plane-fronted gravitational waves is summarised
in the excellent review by Ehlers and Kundt \cite{Ehlers}. The non-linear
pp-waves are exact solutions of Einstein's equations. In the standard 
$(u,v,x,y)$ coordinates, a pp-wave that travels in the $z$ direction is
described by the space-time line element 

\begin{equation}
ds^2=dx^2+dy^2 -2 du\,dv+H(x,y,u)du^2\,.
\label{2}
\end{equation}
These coordinates were first introduced by Brinkmann \cite{Brinkmann}.
The line element depends only on the function $H(x,y,u)$, that must satisfy

\begin{equation}
\nabla^2 H=\biggl( {{\partial^2} \over{\partial x^2}}+ 
{{\partial^2} \over{\partial y^2}} \biggr)H=0\,.
\label{3}
\end{equation}
The dependence of $H$ on the variable $u$ is arbitrary, a property that is 
typical of wave solutions. We will assume that this
dependence is given by a Gaussian function, so that in regions sufficiently far
from the propagating Gaussian the space-time is flat. In these regions we 
usually identify

\begin{equation}
u = {1\over {\sqrt{2}}}(t-z)\,,
\label{4}
\end{equation}

\begin{equation}
v = {1\over {\sqrt{2}}}(t+z)\,.
\label{5}
\end{equation}
In the flat space-time limit, $(t,x,y,z)$ are identified with the standard
Cartesian coordinates.
Unlike our previous works \cite{Artigo1,Artigo2,Artigo3}, we are now adopting
the notation determined by Eq. (\ref{4}). Transforming the line element to the
$(u,x,y,z)$ coordinates, we obtain

\begin{equation}
ds^{2}=(H-2)du^{2}+dx^{2}+dy^{2}-2\sqrt{2}dudz\,.
\label{6}
\end{equation}
This is the form of the line element that will be relevant to the present 
analysis. The establishment of frames is more intuitively 
understood if we 
use spacelike coordinates such as $(x^1,x^2,x^3)=(x,y,z)$ in the line element.
In addition, the use of the coordinates $(u,x,y,z)$ allows to maintain the 
usual interpretation of three-dimensional velocity of particles, which in turn
allows to verify the 
work-energy relation. In view of Eq. (\ref{6}), we are assuming $H<2$. This
condition must be satisfied for the reference frames below.

Now we turn to the construction of a frame adapted to a static observer in
space-time. A given set of tetrad fields $e^a\,_\mu$ yields the space-time
metric tensor $g_{\mu\nu}$ by means of the usual relation 
$e^a\,_\mu e^b\,_\nu \eta_{ab}=g_{\mu\nu}$, where $\eta_{ab}=(-1,+1,+1,+1)$ is
the flat, tangent space metric tensor. The time and space components of the 
indices are denoted by $a=((0),(i)),\; \mu =(0,i)$, where $i=1,2,3$.
In what follows, we will make the speed of light $c=1$.

The inverse tetrads $e_a\,^\mu$ define a frame adapted to a particular class of
observers in space-time. Let the curve $x^\mu(\tau)$ represent the timelike 
worldline $C$ of an observer, where $\tau$ is the proper time of the observer.
The velocity of the observer along $C$ is given by $U^\mu=dx^\mu/d\tau$.
A frame adapted to this observer is constructed by identifying the
timelike component of the frame $e_{(0)}\,^\mu$ with the velocity $U^\mu$
of the observer: $e_{(0)}\,^\mu=U^\mu(\tau)$. 

A static observer in space-time is defined by the condition $U^\mu=(U^0,0,0,0)$
everywhere in the three-dimensional spacelike hypersurface.
Thus, a frame adapted to a static observer in space-time must satisfy the
conditions
$e_{(0)}\,^i=(0,0,0)=e^{(j)}\,_0$. 
In order to characterise the congruence of timelike curves 
$e_{(0)}\,^\mu(x^0,x^i)=(e_{(0)}\,^0,0,0,0)$ that defines a static frame, one 
must fix a time parameter $x^0$, which is the time coordinate for 
all observers in the three-dimensional spacelike hypersurface. In the present 
case of gravitational waves described by a propagating Gaussian, there are flat
regions of the space-time, located far from the wave front.
Once the time coordinate is fixed, the condition $e_{(0)}\,^j(x^0,x^i)=(0,0,0)$ 
holds for any set of spacelike coordinates, since for the transformed
coordinates ${x^\prime}^i={x^\prime}^i(x^j)$ we have 

$${e^\prime}_{(0)}\,^i(x^0,{x^\prime}^k)=
{{\partial {x^\prime}^i}\over {\partial x^j}}\, e_{(0)}\,^j(x^0,x^k)=
(0,0,0)\,.$$
Thus, the static condition on the frame does not depend on the spacelike 
coordinates $x^i$, but depends on the choice of a timelike parameter
$x^0$, that defines the foliation of the space-time in spacelike hypersurfaces.
The condition for static frames in flat space-time regions, in Cartesian 
coordinates, is a very clear and unambiguous concept. It must be noted that when
we assign initial values for the velocities of particles in the flat space-time
(before the passage of the wave, as is done in this article), we are implicitly
assuming the existence of a static frame.

The set of tetrad fields constructed out of the metric tensor (\ref{6}), and 
that is adapted to static observers in space-time is given by 

\begin{eqnarray}
e_{a\mu}(u,x,y,z)=\pmatrix{-A&0&0&-B\cr
0&1&0&0\cr
0&0&1 &0\cr
0&0&0&B\cr}\,.
\label{7}
\end{eqnarray}
where $a$ and $\mu$ denote rows and columns, respectively, 

\begin{equation}
A=\sqrt{2-H}\,, \ \ \ \ B={\sqrt{2} \over {\sqrt{2-H}}}\,,
\label{8}
\end{equation}
and $e=\det (e^a\,_\mu)=\sqrt{2}$. Note that the time parameter in Eq. (\ref{9})
is $x^0=u$, and that the flat space-time tetrad fields in the $(u,x,y,z)$ 
coordinates are 

$$
e^a\,_\mu=\pmatrix{\sqrt{2}&0&0&1\cr
0&1&0&0\cr
0&0&1 &0\cr
0&0&0&1\cr}\,.
$$
In these coordinates, a static observer in space-time is also represented by 
the condition $U^\mu=e_{(0)}\,^\mu=(U^0,0,0,0)$,
which can be verified
by means of a simple coordinate transformation defined by Eq. (\ref{4}).

The acceleration of the observer along an arbitrary worldline $C$ is 
defined by the covariant derivative of $U^\mu=e_{(0)}\,^\mu$ along the
timelike worldline $C$,

\begin{equation}
a^\mu= {{D U^\mu}\over{d\tau}} ={{De_{(0)}\,^\mu}\over {d\tau}} =
\,U^\alpha \nabla_\alpha e_{(0)}\,^\mu\,, 
\label{9}
\end{equation}
where $\tau$ is the proper time of the observer along $C$, and 
the covariant derivative is constructed out of the Christoffel symbols.
Thus, $e_a\,^\mu$ yields the velocity and acceleration of an observer along 
the worldline. Therefore, a given set of tetrad fields, for which 
$e_{(0)}\,^\mu$ describes a congruence of timelike curves, is adapted to a 
particular class of observers, namely, to observers characterized by the 
velocity field $U^\mu=\,e_{(0)}\,^\mu$, endowed with acceleration $a^\mu$. If
$e^a\,_\mu \rightarrow \delta^a_\mu$ in the limit 
$r \rightarrow \infty$, then $e^a\,_\mu$ is adapted to static
observers at spacelike infinity. 

An alternative characterization of tetrad fields as an observer's frame may 
be given by considering the acceleration of the whole frame along an 
arbitrary path $x^\mu(\tau)$ of the observer. The acceleration 
of the whole frame is determined by the absolute derivative of $e_a\,^\mu$ 
along $x^\mu(\tau)$. Thus, assuming that the observer carries an orthonormal 
tetrad  frame $e_a\,^\mu$, the acceleration of the frame along the path is 
given by \cite{Mashh2,Mashh3,Maluf81,Maluf82}

\begin{equation}
{{D e_a\,^\mu} \over {d\tau}}=\phi_a\,^b\,e_b\,^\mu\,,
\label{10}
\end{equation}
where $\phi_{ab}$ is the antisymmetric acceleration tensor. As discussed in 
Refs. \cite{Mashh2,Mashh3}, in analogy with the Faraday tensor we may identify 
$\phi_{ab} \leftrightarrow\ ({\bf a}, {\bf \Omega})$, where ${\bf a}$ is the 
translational acceleration ($\phi_{(0)(i)}=a_{(i)}$) and ${\bf \Omega}$ is the 
frequency of rotation of the local spatial frame  with respect to a 
non-rotating, Fermi-Walker transported frame. 
It follows from Eq. (\ref{10}) that

\begin{equation}
\phi_a\,^b= e^b\,_\mu {{D e_a\,^\mu} \over {d\tau}}=
e^b\,_\mu \,U^\lambda\nabla_\lambda e_a\,^\mu\,.
\label{11}
\end{equation}

The acceleration vector $a^\mu$ may be projected on a frame in order to yield

\begin{equation}
a^b= e^b\,_\mu a^\mu=e^b\,_\mu U^\alpha \nabla_\alpha
e_{(0)}\,^\mu=\phi_{(0)}\,^b\,.
\label{12}
\end{equation}
Thus, $a^\mu$ and $\phi_{(0)(i)}$ are not different translational 
accelerations of the frame. The expression of $a^\mu$ given by Eq. (\ref{9}) 
may be rewritten as

\begin{eqnarray}
a^\mu&=& U^\alpha \nabla_\alpha e_{(0)}\,^\mu 
=U^\alpha \nabla_\alpha U^\mu =
{{dx^\alpha}\over {d\tau}}\biggl(
{{\partial U^\mu}\over{\partial x^\alpha}}
+\,\,^0\Gamma^\mu_{\alpha\beta}U^\beta \biggr) \nonumber \\
&=&{{d^2 x^\mu}\over {d\tau^2}}+\,\,^0\Gamma^\mu_{\alpha\beta}
{{dx^\alpha}\over{d\tau}} {{dx^\beta}\over{d\tau}}\,,
\label{13}
\end{eqnarray}
where $\,\,^0\Gamma^\mu_{\alpha\beta}$ are the Christoffel symbols.
We see that if $U^\mu=\,e_{(0)}\,^\mu$ represents a geodesic
trajectory, then the frame is in free fall and 
$a^\mu=\phi_{(0)(i)}=0$. Therefore we conclude that non-vanishing
values of the latter quantities represent inertial accelerations
of the frame.

An alternative expression of the acceleration tensor is given by
\cite{Maluf81,Maluf82}

\begin{equation}
\phi_{ab}={1\over 2}\lbrack T_{(0)ab}+T_{a(0)b}-T_{b(0)a}\rbrack\,.
\label{14}
\end{equation}
where 

\begin{equation}
T_{abc}=e_b\,^\mu e_c\,^\nu T_{a\mu\nu}=
e_b\,^\mu e_c\,^\nu (\partial_ \mu e_{a\nu}-\partial_\nu e_{a\mu})\,.
\label{15}
\end{equation}
The tensor $\phi_{ab}$ is invariant under coordinate transformations and 
covariant under global SO(3,1) transformations, but not under
local SO(3,1) transformations. Because of this property, $\phi_{ab}$ may be
used to characterise the inertial state of the frame. If the frame is 
maintained static in space-time, then the six components of the tensor 
$\phi_{ab}$ cancel the six components of the gravitational acceleration. 

\section{The gravitational acceleration on particles and the work-energy
relation in the pp-wave space-time}

The acceleration tensor in the form given by Eq. (\ref{14}) is more useful for
practical purposes. For the set of tetrad fields given by Eq. (\ref{7}), the
non-vanishing components of $\phi_{ab}$ are

\begin{eqnarray}
\phi_{(0)(1)}&=&\partial_{x}\ln{A}=-\frac{1}{2}\,\frac{\partial_{x}H}{2-H}\,,
\label{16} \\
\phi_{(0)(2)}&=&\partial_{y}\ln{A}=-\frac{1}{2}\,\frac{\partial_{y}H}{2-H}\,,
\label{17} \\
\phi_{(0)(3)}&=&-\partial_{u}\left(\frac{1}{A}\right)=
-\frac{1}{2}\,\frac{\partial_{u}H}{(2-H)^{3/2}}\,,
\label{18} \\
\phi_{(1)(3)}&=&-\frac{1}{2}\,\frac{\partial_{x}H}{2-H}=\phi_{(0)(1)}\,,
\label{19} \\
\phi_{(2)(3)}&=&-\frac{1}{2}\,\frac{\partial_{y}H}{2-H}=\phi_{(0)(2)}\,.
\label{20} \\ \nonumber
\end{eqnarray}
The tetrad field components are valid in the space-time region where
$A=\sqrt{2-H}$ is well defined, namely, where $(2-H)>0$. For this reason,
considering the various possibilities for the function $H(u,x,y)$, we choose the
gravitational wave profile of the Aichelburg-Sexl solution for a non-linear
gravitational wave \cite{Aichelburg}, since it yields a tetrad frame defined 
almost everywhere in the three-dimensional space. The Aichelburg-Sexl solution
for the function $H$, with a suitable multiplicative sign, reads

\begin{equation}
H=-{1\over 2} \ln \biggl( {\rho \over {\rho_0}}\biggr)e^{-u^2}\,,
\label{21}
\end{equation}
where $\rho=\sqrt{x^2+y^2}$, and $\rho_0$ is an arbitrary constant with 
dimension of length. The condition $(2-H)>0$ is satisfied if

\begin{equation}
\rho > \rho_0 e^{-4\,e^{(u^2)}}\,.
\label{22}
\end{equation}
The constant $\rho_0$ can be chosen to be arbitrarily small. However, in the 
analysis below, we will consider for simplicity $\rho_0=1$ in natural units.
The right hand side of the equation above vanishes for $u=\pm \infty$. 
We will restrict considerations to the vacuum regions of the space-time.
Considering Eq. (\ref{21}), we find

\begin{eqnarray}
\phi_{(0)(1)}&=&\frac{\cos{\phi}}{\rho\left(8e^{u^{2}}+2\ln{\rho}\right)}\,,
\label{23} \\
\phi_{(0)(2)}&=&\frac{\sin{\phi}}{\rho\left(8e^{u^{2}}+2\ln{\rho}\right)}\,,
\label{24} \\
\phi_{(0)(3)}&=&-\frac{4ue^{-u^{2}}\ln{\rho}}{\left(8+ 2 e^{-u^{2}}
\ln{\rho}\right)^{3/2}}\,.\label{25} \\
\nonumber
\end{eqnarray}

The gravitational acceleration of the particles that are hit by the wave is
taken to be exactly minus the inertial acceleration that is necessary to keep
the frame static everywhere in the three-dimensional space. This is the 
condition that establishes the static frame in the present analysis. Thus, the
components above will be multiplied by $-1$ in order to yield the 
gravitational acceleration ${\bf a}_g$, 

\begin{eqnarray}
{\bf a}_g&=&-\frac{1}{\rho\left(8e^{u^{2}}+ 2\ln{\rho}\right)}
\hat{\rho} +\frac{4ue^{-u^{2}}\ln{\rho}}
{\left(8 + 2 e^{-u^{2}}\ln{\rho}\right)^{3/2}}\hat{z} \nonumber \\
&=& -\frac{1}{4\rho}\frac{e^{-u^{2}}}
{A^{2}}\hat{\rho} + u\frac{H}{A^{3}}\hat{z} \nonumber \\
& \equiv& a_{g\rho}\hat{\rho}+ a_{gz}\hat{z}\,,
\label{26}
\end{eqnarray}
where $\hat{\rho}=\cos\phi\,\hat{x}+\sin\phi\,\hat{y}$.

Since $a_{g\rho}<0$, $a_{g\rho}$ is always 
attractive, i.e., it is oriented towards the axis $\rho=0$. Had we considered
the $+$ sign on the right hand side of Eq. (\ref{21}), $a_{g\rho}$ would be 
always repulsive.

Now we turn to the evaluation of the work per unit mass $\Delta W$ done by the
gravitational field on the particles, making use of the ordinary Newtonian 
concepts. Let $i$ represent the initial position of
the particle, when the wave starts to act upon the particle, and $f$ the final
position, when the action of the wave upon the particle is ceased
(i.e., the positions when the wave is 
approaching the particle and when the wave is leaving behind the particle,
respectively). Far from the wave, namely, when the gravitational field of the
wave does not affect the particle, the geodesic trajectory of the particle is a
straight line, since $\phi_{(0)(i)}=0$ in such regions of the three-dimensional
space. The fall-off of the gravitational field outside of the ``sandwich'' 
region is guaranteed by the Gaussian function in Eq. (\ref{21}).
The work per unit mass done by the gravitational field is given by

\begin{equation}
\Delta W=-\int_i^f {\phi_{(0)(i)}e^{(i)}\,_{j} dx^{j}}=-\int_i^f 
a_j dx^j\,,
\label{27}
\end{equation}
with $\phi_{(0)(i)}$ given by Eqs. (\ref{23}-\ref{25}). The second equality
is obtained from Eq. (\ref{12}), which may be inverted to yield
$a^\mu=\phi_{(0)(i)}e^{(i)\mu}$, where $\lbrace a^j \rbrace$ are the spatial 
components of the inertial acceleration that maintain the frame static in
space-time, and are given by Eq. (\ref{13}). We note the similarity of the right
hand side of Eq. (\ref{27}) with the right hand side of Eq. (\ref{1}). The 
limits $i$ and $f$ represent asymptotic states, but in the numerical 
calculations it suffices to take some finite, large value of $u$, such as in the
initial conditions (\ref{34}-\ref{35}) below. The expression above becomes

\begin{eqnarray}
\Delta W &=& -\int_i^f{\left[\partial_{x}\ln{(A)}\,dx+\partial_{y}\ln{(A)}\,dy-
\frac{1}{\sqrt{2}}\partial_{u}\left(\frac{1}{A^2}\right)dz\right]} \nonumber \\
&=& -\int_{-\infty}^{\infty}{\left[\partial_{x}\ln{(A)}\,\dot{x}+
\partial_{y}\ln{(A)}\, \dot{y}-\frac{1}{\sqrt{2}}\partial_{u}\left(\frac{1}{A^2}
\right)\dot{z}\right]du}\,,
\label{28}
\end{eqnarray}
where we have used $dx^{j}=(dx^{j}/{du})du=\dot{x}^{j}du$. By substituting the 
expression of $A$, we find

\begin{eqnarray}
\Delta W&=& -\int_{-\infty}^{+\infty} \biggl[ 
{{x\,\dot x}\over {4(x^2+y^2)(2-H)}}+{{y\,\dot y}\over {4(x^2+y^2)(2-H)}} 
\nonumber \\
&{}&-{{\sqrt{2} e^{-u^2}\,u\,\ln(x^2+y^2)\dot{z}}\over {4(2-H)^2}}\biggr]du\,.
\label{29}
\end{eqnarray}
This integral can be solved numerically by taking into account  the solutions 
$x(u)$, $y(u)$ and $z(u)$, which can be obtained numerically by using the
program MATHEMATICA. These solutions are obtained from the Eqs. 
(\ref{31}), (\ref{32}) and (\ref{33}) below.

The variation $\Delta K$ of the kinetic energy of the particles is obtained
from the expression

\begin{eqnarray}
\Delta K &=&\frac{1}{4}\biggl[ \dot{x}^2(u)+\dot{y}^2(u)
+\dot{z}^2(u) \biggr]_{u=-\infty}^{u=\infty} \nonumber \\
&=&\frac{1}{2}\biggl[ \biggl( {{dx}\over{dt}}\biggr)^2 +
\biggl( {{dy}\over{dt}}\biggr)^2+
\biggl( {{dz}\over{dt}}\biggr)^2 \biggr]_{t=-\infty}^{t=\infty}\,.
\label{30}
\end{eqnarray}
Considering that the whole work done by the gravitational field on the particle
is converted into kinetic energy, we expect $\Delta K \approx \Delta W $. 
We are clearly assuming a non-relativistic behaviour of the particle, which is
a realistic situation in the context of gravitational wave measurements, 
in which case the gravitational field is very weak. 

Equation (\ref{30}) for
$\Delta K$ represents the standard classical expression of kinetic energy for 
static observers in space-time, as defined in the previous section, in the 
context of non-relativistic particles (for relativistic particles, 
see Eq. (\ref{53}) below). Since the definition (\ref{27}) of the work per unit 
mass $\Delta W$ explicitly depends on the set of tetrad fields given by 
Eq. (\ref{7}), then $\Delta W$ is also defined in the frame of static observers
in space-time.

The geodesic equations in the $(u,x,y,z)$ coordinates for the particle motion in
space-time are \cite{Artigo1,Artigo3}

\begin{eqnarray}
\ddot{x}&=&\frac{1}{2}\partial_{x}H \,, \label{31} \\ 
\ddot{y}&=&\frac{1}{2}\partial_{y}H \,, \label{32} \\
\sqrt{2} \ddot{z}&=& \frac{1}{2} \partial_{u}H +
(\partial_{x}H)\dot{x}+(\partial_{y}H)\dot{y} \label{33} \,.\\
\nonumber
\end{eqnarray}
The parameter along the geodesics is $u$. In what follows, we have chosen 
initial conditions and trajectories for the particles such that Eq. (\ref{22})
is satisfied. 

The work-energy relation is verified by considering two sets of initial 
conditions, one in which the initial velocity $\dot{x}(u_0)$ in the $x$ 
direction is unspecified, and the other in which the initial position 
$x(u_0)$ is unspecified. We will take $u_0=-20$ in natural units, and consider

\begin{equation}
\textsl{I}:\, x(-20)=5\,, \   y(-20)=0\,, \    z(-20)=0\,, \   
\dot{y}(-20)=0\,, \   \dot{z}(-20)=0\,,
\label{34}
\end{equation}

\begin{equation}
\textsl{II}:\,  y(-20)=0\, \ z(-20)=0\,,  \ \dot{x}(-20)=0\,, 
\ \dot{y}(-20)=0\,, \  \dot{z}(-20)=0\,.
\label{35}
\end{equation}

In the figures below we, plot altogether $\Delta K$ and $\Delta W$, first
varying the initial velocity $\dot{x}(u_0)\equiv \dot{x}_0$, with initial 
conditions $\textsl{I}$, and then varying the initial position 
$x(u_0)\equiv x_0$, with initial conditions \textsl{II}.

\begin{figure}[H]
	\centering
		\includegraphics[width=0.7\textwidth]{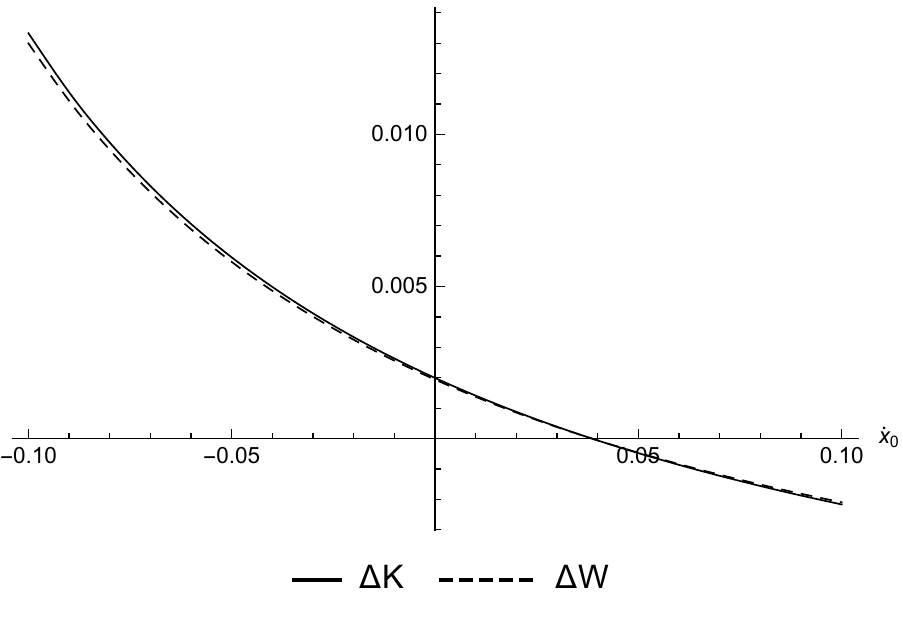}
	\caption{$\Delta K$ (continuous line) and $\Delta W$ (dashed line) for the
		initial conditions I.}
	\label{Figure1}
\end{figure}

\begin{figure}[H]
	\centering
		\includegraphics[width=0.7\textwidth]{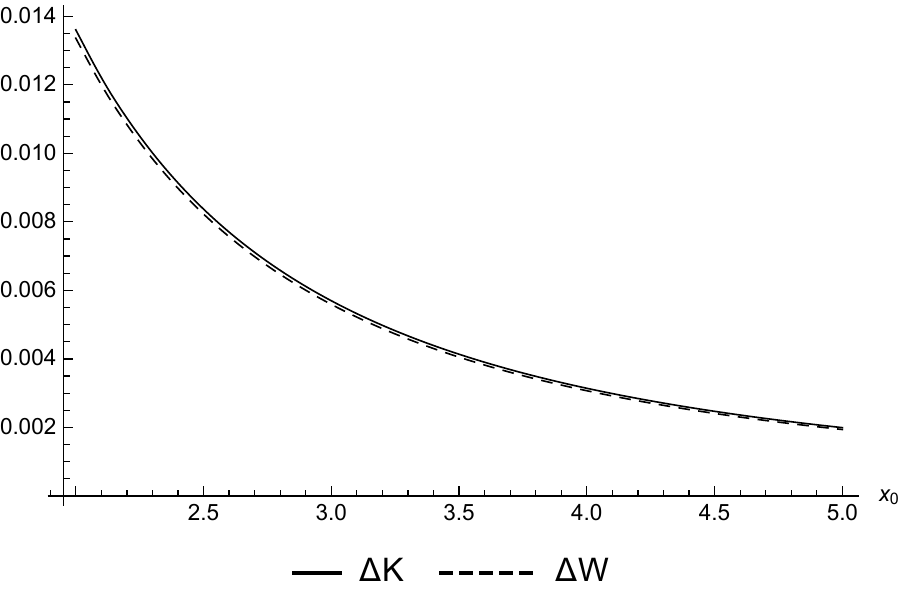}
	\caption{$\Delta K$ (continuous line) and $\Delta W$ (dashed line) for the
			initial conditions II.}
	\label{Figure2}
\end{figure}

For both initial conditions, the difference between $\Delta K$ and 
$\Delta W$ is less than $2$ percent. Part of this difference may be due to the
numerical errors in the evaluation of the quantities. But the figures 
undoubtedly confirm  that the work-energy relation is valid in the 
non-relativistic approximation,
and demonstrate that the difference between the initial and final kinetic
energy of the particle is due to the gravitational acceleration imparted by the
gravitational wave. We believe that there is a close relationship between 
the gravitational acceleration imparted to the particles in the gravitational
wave space-time, and the gravitational centre of mass $M$, which may be thought
as a ``visualisation of the centre of gravity'' of the gravitational wave. For
positive $M$, the gravitational acceleration should be negative (attractive), 
and vice-versa. For this reason, we address the gravitational centre of mass in
the following section.

\section{The gravitational centre of mass in the teleparallel equivalent of 
general relativity}

The teleparallel equivalent of general relativity (TEGR) is an alternative
geometrical description of the relativistic theory of gravitation, in which one
can establish the notion of distant parallelism, as discussed by M\o ller.
In a space-time described by a set of tetrad fields, two  vectors 
at distant points are called parallel \cite{Mol} if they have identical 
components with respect to the local tetrads at the points considered. Let us
consider a vector field $V^\mu(x)$. At the point $x^\lambda$ its tetrad 
components are $V^a(x)=e^a\,_\mu(x)V^\mu(x)$. For the tetrad components 
$V^a(x+dx)$ at $x^\lambda + dx^\lambda$, it is easy to see 
that $V^a(x+dx)=V^a(x)+\nabla V^a(x)$, where 
$\nabla V^a(x)=e^a\,_\mu(\nabla_\lambda V^\mu)dx^\lambda$. The covariant 
derivative $\nabla$ is constructed out of the Weitzenb\"{o}ck connection
$\Gamma^\lambda_{\mu\nu}=e^{a\lambda} \partial_\mu e_{a\nu}$. The vanishing of
the covariant derivative $\nabla_\lambda V^\mu$ establishes the parallelism of
the vector field in space-time. In particular, the tetrad fields are 
auto-parallels, since $\nabla_\lambda e_a\,^\mu\equiv 0$. It is 
interesting to note that the notion of distant parallelism shares a conceptual
similarity with the notion of entanglement in quantum theory. 
But the distant parallelism is lost if one allows for the presence of a
connection that yields local Lorentz covariance of the field quantities.
By performing arbitrary local Lorentz transformations, two distant vectors 
that are initially parallel, will no longer maintain the distant parallelism.
We do not make use of any arbitrary, unspecified (flat or non-flat)
connection.

In the TEGR \cite{Maluf1}, the field equations are covariant under local 
Lorentz transformations, without the necessity of any connection, 
and are precisely equivalent to Einstein's equations. The meaning of this 
covariance is that the theory is valid in any frame in space-time. But field
quantities such as energy, momentum and 4-angular momentum of the 
gravitational field, are not invariant under 
local transformations, but covariant under global Lorentz transformations.
These quantities, in classical or relativistic physics, are frame dependent. 
In particular, the concept of gravitational energy is frame
dependent, and is not identical to notions such as the total mass of a black 
hole space-time, for instance, which is related to the rest mass of the black 
hole. Note that an isolated black hole with velocity $v$, when observed at very
large distances, may be considered as a particle of rest mass $m$, say, with 
energy $E=\gamma mc^2$, where $\gamma = (1-v^2/c^2)^{-1/2}$.

The Hamiltonian formulation of the TEGR is well established (see \cite{Maluf1} 
and references therein). The definitions of energy, momentum and 4-angular
momentum of the gravitational field are obtained in the realm of the Hamiltonian
formulation, but the balance (conservation) equations for the gravitational 
energy-momentum are obtained in the Lagrangian framework. Here, we will just 
recall that these definitions satisfy the algebra of the Poincar\'e group.
This algebra is constructed by calculating the Poisson brackets in the phase 
space of the theory between the gravitational energy-momentum vector $P^a$ 
and the 4-angular momentum $L^{ab}$, and is given by \cite{Maluf83}

\begin{eqnarray}
\lbrace P^a , P^b \rbrace &=& 0\,, \nonumber \\
\lbrace P^a , L^{bc} \rbrace &=& \eta^{ab} P^c-\eta^{ac} P^b 
\,, \nonumber \\
\lbrace L^{ab}, L^{cd} \rbrace &=&
\eta^{ac}L^{bd} +\eta^{bd}L^{ac} -\eta^{ad}L^{bc}-\eta^{bc}L^{ad}\,.
\label{36}
\end{eqnarray}
In the expression of $L^{ab}$ below, we are adopting the convention of Refs.
\cite{Maluf1,Maluf2}, which differs by a minus sign from Ref. \cite{Maluf83}.
$L^{ab}$ are obtained from the primary, first class constraints of the 
Hamiltonian formulation of the theory, and are defined by

\begin{equation}
L^{ab} = -\int_{V}d^{3}x\,M^{[ab]}\,,
\label{37}
\end{equation}
where 
\begin{equation}
M^{[ab]}  
= 2k\partial_{i}[e(e^{ai}e^{b0} - e^{bi}e^{a0})]\,,
\label{38}
\end{equation}
and $k=c^3/(16\pi G)=1/(16\pi)$.
The gravitational centre of mass (COM) components are 

\begin{equation}
L^{(0)(i)}=-\int d^3x\, M^{[(0)(i)]}  \,,
\label{39}
\end{equation}
where

\begin{equation}
M^{[(0)(i)]}=2k\partial_j\lbrack e(e^{(0)j}e^{(i)0}-
e^{(i)j}e^{(0)0})\rbrack\,.
\label{40}
\end{equation}
The quantity $-M^{(0)(i)}$ is identified as the gravitational COM density, and
is determined once we establish the suitable set of tetrad fields that defines 
a frame in space-time, in this case adapted to a static observer in space-time.
The gravitational COM has been investigated in Ref. \cite{Maluf2}, in the 
context of several configurations of the gravitational field.

The set of tetrad fields in consideration are given by Eqs.  (\ref{7}-\ref{8}).
The non-vanishing components of the COM density are

\begin{equation}
-M^{(0)(1)}=-2\sqrt{2}k\partial_{x}\left(\frac{1}{A}\right)=
\frac{k}{\sqrt{2}A^3}\frac{\cos{\phi}}{\rho}e^{-u^2}\,,
\label{41}
\end{equation}

\begin{equation}
-M^{(0)(2)}=-2\sqrt{2}k\partial_{y}\left(\frac{1}{A}\right)=
\frac{k}{\sqrt{2}A^3}\frac{\sin{\phi}}{\rho}e^{-u^2}\,,
\label{42}
\end{equation}
where $\phi$ is the azimuthal angle. These components may be presented in 
vector form according to

\begin{equation}
{\bf{M}}=\frac{k}{\sqrt{2}A^3}\frac{1}{\rho}e^{-u^2}\hat{\rho}\,
\equiv M \hat{\rho}
\label{43}
\end{equation}
In Figure 3 we plot $M$ as a function of $u$ and $\rho$.

\begin{figure}[H]
	\centering
		\includegraphics[width=0.6\textwidth]{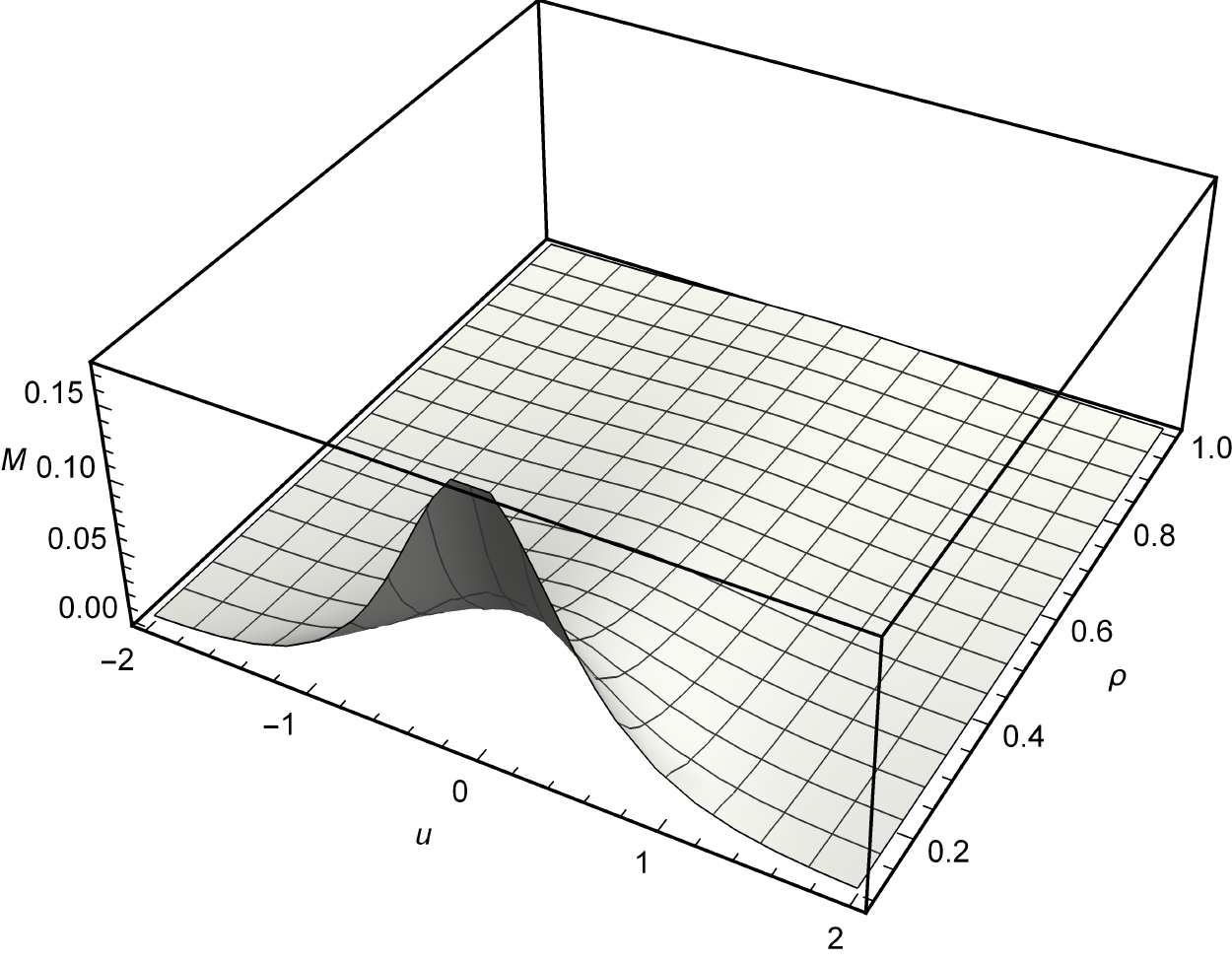}
	\caption{The gravitational COM density $M$ given by Eq. (\ref{43}). The 
		core of the wave burst is concentrated around $u=0$. The gravitational
		COM is similarly more intense around $u=0$, and vanishes when 
		$\rho \rightarrow \infty$ and $u \rightarrow \pm \infty$. Therefore
		it is related to the localization of the wave burst.}
		\label{Figure3}
\end{figure}
As in Ref. \cite{Maluf2}, we see that the radial gravitational acceleration of
a geodesic particle in space-time, given on the right hand side of 
Eq. (\ref{26}) for the wave determined
by Eq. (\ref{21}), is negative and therefore
directed towards the region with higher intensity of the gravitational centre of
mass density, which plays the role of ``centre of gravity''.
The propagation of the gravitational wave 
may also be thought as the propagation in space-time of the COM density shown 
in Figure 3. Had we considered the plus sign in Eq. (\ref{21}), the radial
acceleration would be repulsive, with a minus sign multiplying the right hand
side of Eq. (\ref{26}), and also the right hand side of Eqs. (\ref{41}-\ref{43})
would be multiplied by $-1$.

\section{Oscillating pp-wave}

In this section we will consider the interesting case of an oscillating pp-wave.
We will just multiply the Gaussian function by a harmonic function in the
expression of $H(u,\rho)$, and again in this case the 
work-energy relation is verified. Let us consider the function

\begin{equation}
H=-\frac{1}{2}\ln{(\rho)}\cos{(u)}e^{-(u/10)^2}\,,
\label{44}
\end{equation}
in the frame defined by Eqs. (\ref{7}-\ref{8}). The resulting inertial 
accelerations are

\begin{eqnarray}
\phi_{(0)(1)}&=&\frac{\cos{\phi}\cos{u}}{4\rho\left(2-H\right)}e^{-(u/10)^2}
=\phi_{(1)(3)}\,, 
\label{45} \\
\phi_{(0)(2)}&=&\frac{\sin{\phi}\cos{u}}{4\rho\left(2-H\right)}e^{-(u/10)^2}
=\phi_{(2)(3)}\,, 
\label{46} \\
\phi_{(0)(3)}&=&-\frac{\left(\frac{u\cos{u}}{50}+\sin{u}\right)\ln{\rho}}
{4\left(2-H\right)^{3/2}}e^{-(u/10)^2}\,.
\label{47} \\
\nonumber
\end{eqnarray}
As before, the gravitational acceleration on the frame is exactly minus the 
expressions above. In vector form, the gravitational acceleration is written as

\begin{eqnarray}
{\bf{a}}_{g}&=&-\frac{\cos{u}}{4\rho\left(2-H\right)}e^{-(u/10)^2}\hat{\rho}
+\frac{\left(\frac{u\cos{u}}{50}
+\sin{u}\right)\ln{\rho}}{4\left(2-H\right)^{3/2}}e^{-(u/10)^2}\hat{z} 
\nonumber \\
&\equiv& a_{g\rho}\hat{\rho}+ a_{gz}\hat{z}\,. \label{48} \\
\nonumber
\end{eqnarray}
This is the acceleration of the particles that are hit by the wave.

The radial and longitudinal components of the gravitational acceleration above
are displayed in Figures 4 and 5, respectively. 

\begin{figure}[H]
	\centering
		\includegraphics[width=0.6\textwidth]{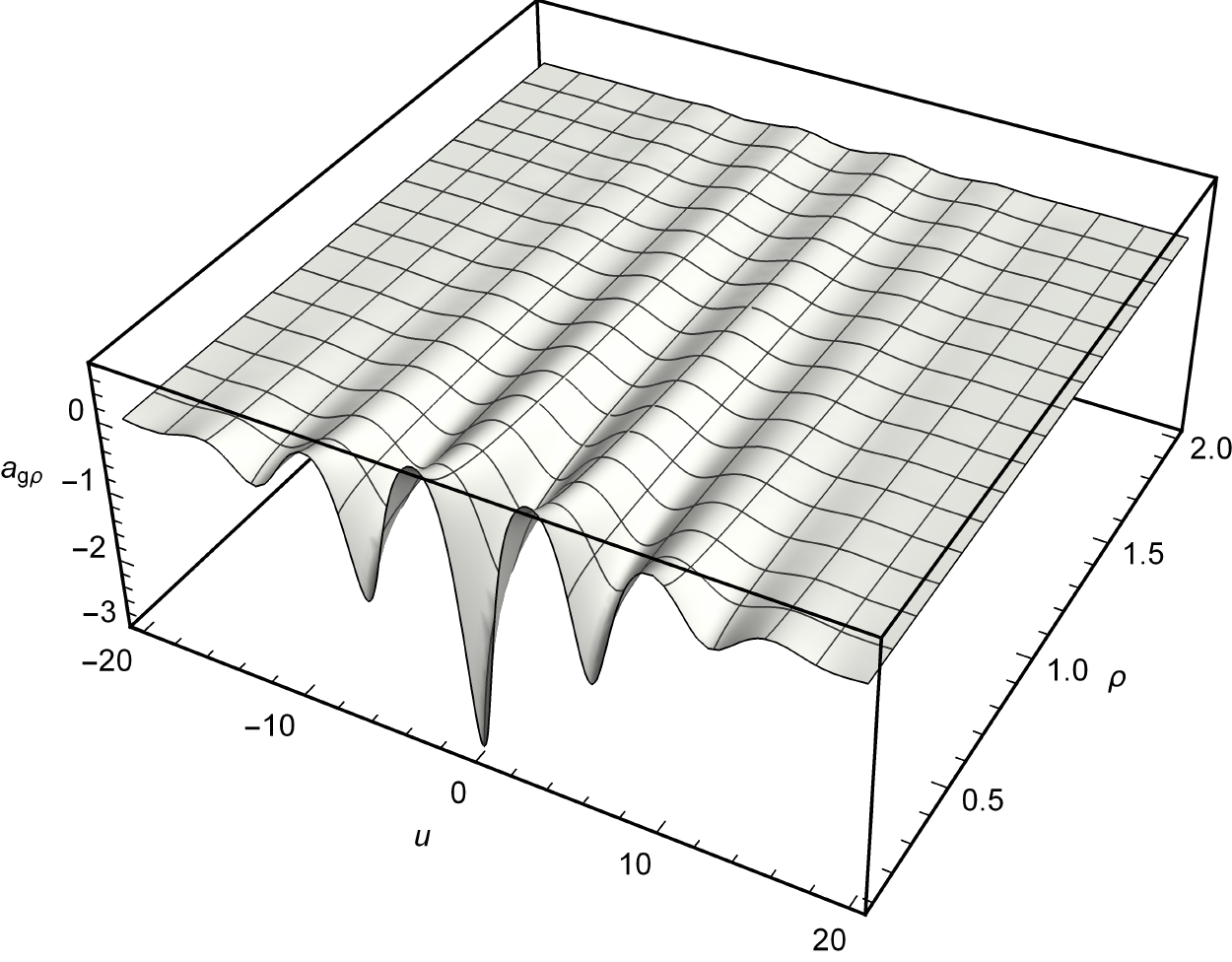}
	\caption{Radial gravitational acceleration for $H$ given 
		by (\ref{44}).}
	\label{Figure4}
\end{figure}

\begin{figure}[H]
	\centering
		\includegraphics[width=0.6\textwidth]{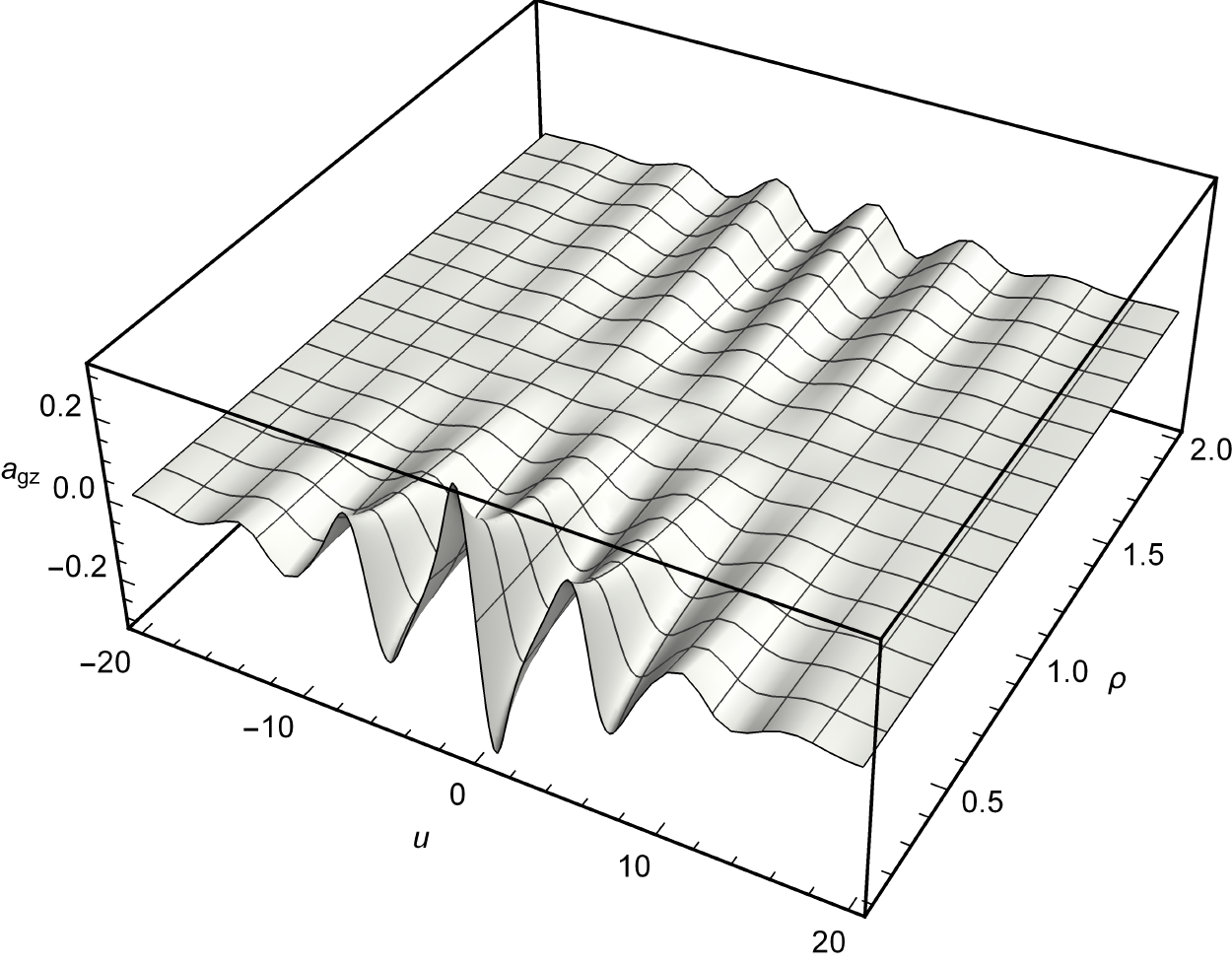}
	\caption{Longitudinal gravitational acceleration for $H$ 
	given by (\ref{44}).}
	\label{Figure5}
\end{figure}

The variations of the longitudinal acceleration, as given in Figure 5, suggests
a trapping of a particle, but only along the $z$ direction. To some extent, this
feature is similar to the trapping of particles discussed in Ref. \cite{ZDGH4}. 

We also analyse the distribution of the COM density for the 
oscillating wave. The non-vanishing components of the COM density are

\begin{eqnarray}
-M^{(0)(1)}&=&\frac{k\cos{u}}{\sqrt{2} A^{3}}\frac{\cos{\phi}}
{\rho}e^{-(u/10)^2}\,, 
\label{49} \\
-M^{(0)(2)}&=&\frac{k\cos{u}}{\sqrt{2} A^{3}}\frac{\sin{\phi}}
{\rho}e^{-(u/10)^2}\,,
\label{50} \\
\nonumber
\end{eqnarray}
where $A=\sqrt{2-H}$. In vector form, we have

\begin{equation}
{\bf{M}}=\frac{k\cos{u}}{\sqrt{2} A^{3}}\;\frac{1}{\rho}\;
e^{-(u/10)^2}\hat{\rho} \equiv M \hat{\rho}\,.
\label{51}
\end{equation}

In Figure 6 we plot the quantity $M$ defined above with respect to both 
$u$ and $\rho$. Regions with higher, positive values of $M$ are attractive. It
is interesting to note that Figures 4 and 6 are very much similar, except that
one is inverted with respect to the other, and the same feature takes place
in the former case of a non-oscillating wave. We see a relationship between the
radial acceleration $a_{\rho}$ and the quantity $M$
defined by Eq. (\ref{51}), which is given by

\begin{equation}
a_{\rho}=-(4\sqrt{2}\pi\,A)\,M\,.
\label{52}
\end{equation}
Thus, if $M>0$, then $a_{\rho}$ is attractive. This feature justifies the 
consideration of the COM density of the waves. There is a not yet totally 
understood relationship between the gravitational acceleration and the 
gravitational COM density, in general.

\begin{figure}[H]
	\centering
		\includegraphics[width=0.7\textwidth]{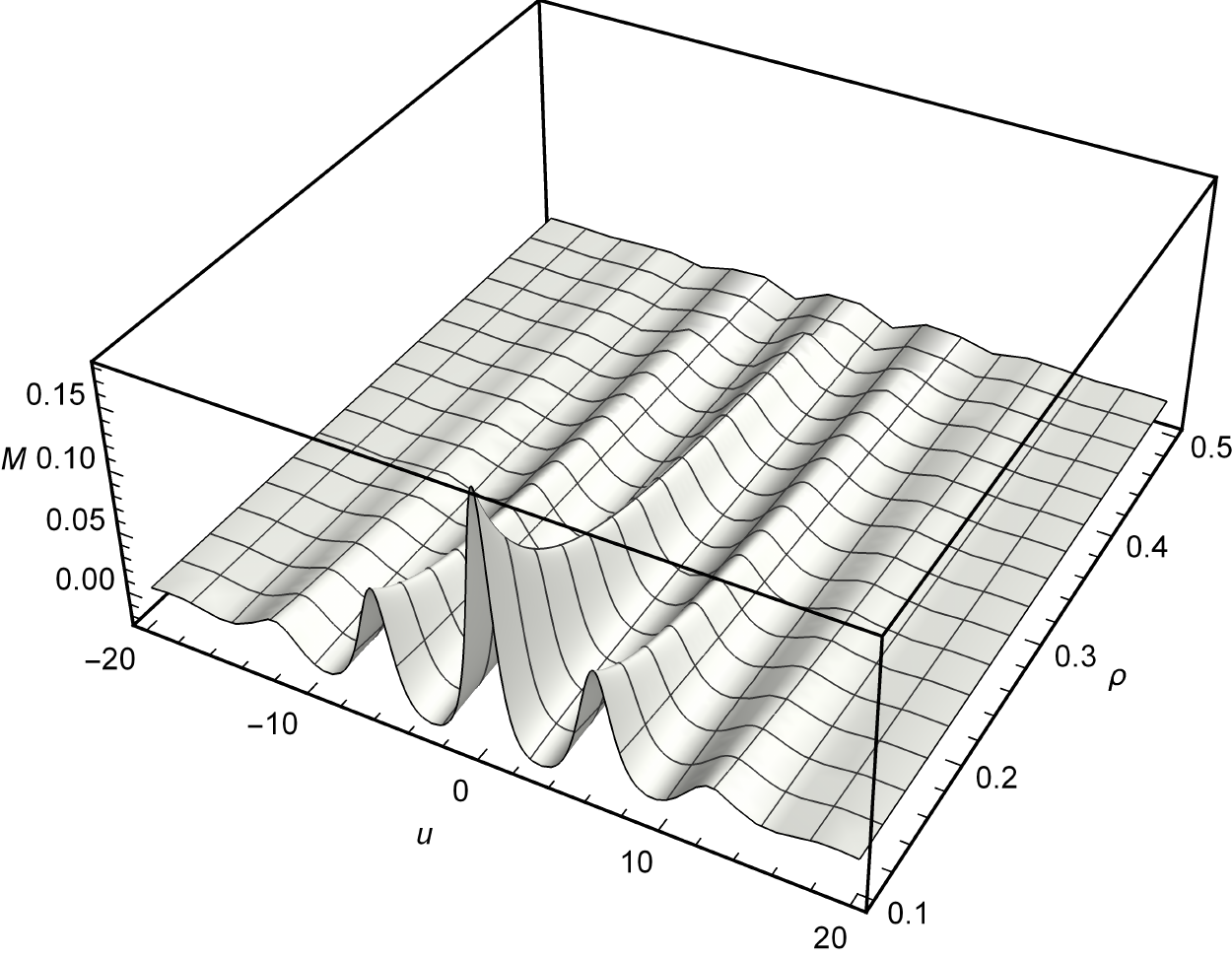}
	\caption{Centre of mass for $H$ given by (\ref{44}).}
	\label{Figure6}
\end{figure}

Finally, we analyse the validity of the work-energy relation for 
particles on geodesics in the oscillating pp-wave determined by Eq. 
(\ref{44}). Considering the initial conditions \textsl{II}, given by
Eq. (\ref{35}), where the initial condition $x(u_0)=x_0$ may be varied,
we find the results displayed in Figure 7.

\begin{figure}[H]
	\centering
		\includegraphics[width=0.7\textwidth]{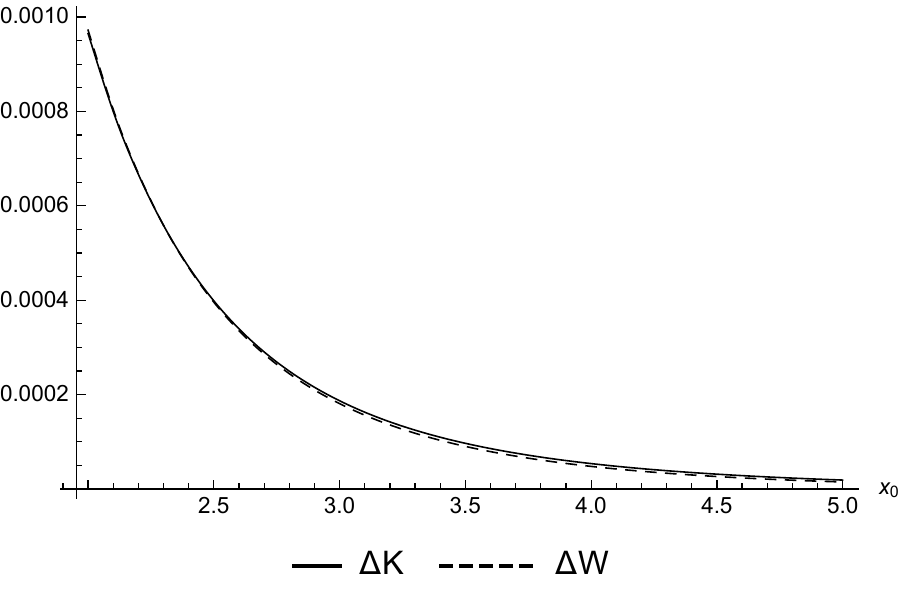}
	\caption{$\Delta K$ (continuous line) and $\Delta W$ (dashed line) for the
			initial conditions \textsl{II}, and $H$ given by Eq. (\ref{44}).}
	\label{Figure7}
\end{figure}
Once again, we verify that the work-energy relation is valid to an excellent  
approximation in the context of the oscillating pp-wave, which shows that the 
particles are indeed subject to the gravitational acceleration due to the
wave.

\section{Summary and conclusions}

In this article we have considered the action on free particles of non-linear
gravitational pp-waves, that propagate along the $z$ axis. The geodesic 
behaviour of these particles is such that before and 
after the passage of the wave, the particle trajectories are straight lines, 
i.e., they are free. We have carried out an attempt to explain why the 
final kinetic energy of particles may be smaller or larger than the initial 
values. The attempt is based on the classical work-kinetic energy relation of 
Newtonian physics $\Delta K=\Delta W$, where the work 
$\Delta W$ is given by Eq. (\ref{27}), and 
$\Delta K$ by Eq. (\ref{30}). Both $\Delta W$ and $\Delta K$ are quantities 
per unit mass. Figures 1, 2 and 7 show that the values of $\Delta W$ and
$\Delta K$ are almost the same, by varying some of the initial conditions. That
these two quantities are practically the same is a surprise, because $\Delta W$
is a relativistic quantity, whereas $\Delta K$ is clearly non relativistic.
The difference between $\Delta W$ and $\Delta K$ may be due to this
fact, as well as to some errors in the numerical (computer) evaluation of the
quantities. In view of these results, we are led to define Eq. (\ref{27}) as 
an exact, relativistic expression for the variation of the kinetic energy per 
unit mass $\Delta\textsf{K}$ of the particles. Thus, we define

\begin{equation}
\Delta \textsf{K}=\Delta W=-\int_i^f {\phi_{(0)(i)}e^{(i)}\,_{j} dx^{j}}\,,
\label{53}
\end{equation}
for particles along geodesic trajectories. This expression is well defined as
long as the notion of kinetic energy makes sense for geodesic 
particles. The equation above may be used to compute the variation of the
kinetic energy of relativistic particles that travel at velocities close to the
velocity of light, in the pp-wave space-times.

In the context of the gravitational waves considered in this article, there is
an interesting relation between the radial acceleration $a_{\rho}$ and the
intensity $M$ of the COM density. This relation is given by Eq. (\ref{52}).
Positive or negative values of $M$ imply attractive or repulsive radial
gravitational acceleration, respectively. This relation will be investigated
elsewhere, in the consideration of general gravitational field configurations.

One important conclusion of our analysis is that non-linear gravitational
waves impart accelerations to particles that are hit by the wave. This 
acceleration, together with the initial conditions of the particle, lead to an
increase or decrease of the energy of the particle, which implies, respectively,
a decrease or increase of the gravitational energy of the wave, since the 
gravitational field is doing work on the particle. The quantity $\Delta W$ 
defined by Eqs. (\ref{27}) and (\ref{53}) determines the energy per unit mass 
that is exchanged between the particle and the gravitational wave. If the 
gravitational field does a positive work on the particle, the energy of the 
latter is increased, and a negative work implies $\Delta \textsf{K}<0$, exactly
as in Newtonian mechanics. We expect 
Eq. (\ref{1}) to be valid also for a charged particle if ${\bf a}$ is, in
this case, the acceleration due to an electromagnetic wave, for instance. 
Therefore Eq. (\ref{1}), understood as a generalised Torricelli equation,  may
have an universal character. It could be possible that the issues analysed in
this article are also described in the context of the Newton-Cartan 
formulation of gravity.

\end{document}